\documentstyle{mn}
\begin{document}
\def\simlt{\mathrel{\rlap{\lower 3pt\hbox{$\sim$}}
        \raise 2.0pt\hbox{$<$}}}
\def\simgt{\mathrel{\rlap{\lower 3pt\hbox{$\sim$}}
        \raise 2.0pt\hbox{$>$}}}
\title[Optical identifications of $\sim 4000$ radio sources at the 1 mJy level]
{Optical identifications of ${\bf\sim 4000}$ radio sources at the 1 mJy level}
\author[M. Magliocchetti, S. J. Maddox]
{M.~Magliocchetti $^{1}$,
S. J. Maddox ${^2}$\\
$^1$SISSA, Via Beirut 4, 34100, Trieste, Italy \\
$^2$School of Physics and Astronomy, University of Nottingham,
Nottingham NG7 2RD, UK}

\maketitle
\vspace {7cm }

\begin{abstract}  

We present an analysis of the properties of optical counterparts
of radio sources down to 1~mJy.
Optical identifications have been obtained by matching together
objects from the APM and FIRST surveys over the region
$148.13^\circ\simlt {\rm RA(2000)}\simlt 218.13^\circ$,
$-2.77^\circ\simlt {\rm Dec} \simlt 2.25^\circ$.  Selecting radio
sources down to 1~mJy, and adopting a uniform optical limit of $b_J =
21.5$ we find 3176 have a counterpart in the APM catalogue,
corresponding to 13 per cent of the radio sample.

For $b_J \le 20.5$ we can divide radio sources into resolved radio
galaxies and stellar-like objects (principally QSO). We find the
population of radio galaxies to be mainly made by early-type galaxies
with very red colours ($b_J-R$ up to $\sim 4$) and a radio-to-optical
ratio $10^2\simlt r\simlt 10^4$.  The contribution of starbursting
objects is negligible. In general QSO show $r\simgt 10^4$ and
$0\simlt b_J -R \simlt 1$.  On the basis of the $R$ magnitudes, we
estimate the sample of radio galaxies to be complete up to $z\simeq
0.3$. We can therefore divide the whole sample of radio sources into a
low-z and high-z population. The low-z one includes the objects
identified as galaxies in the APM survey, the high-z one includes 
sources either identified as QSO or with no optical counterpart for $b_J
\le 20.5$. 

We find that radio galaxies are strongly clustered and highly biased
tracers of the underlying mass distribution.  Models for the angular
correlation function $w(\theta)$ show good agreement with the
observations if we assume a bias factor $b\simeq 2$ at $z\simeq 0.3$.

\end{abstract}


\section{INTRODUCTION}
The advent of a new generation of large area radio surveys which
sample the radio sky down to mJy levels (e.g. FIRST, Becker et al.,
1995; NVSS, Condon et al., 1998; SUMSS, Bock et al., 1999) has
introduced great advantages for both radio astronomy and
cosmology. From the cosmological point of view, the high surface
density of sources allows studies of the large-scale structure of the
Universe and its evolution up to high redshifts ($z\sim 4$) and large
physical scales (see e.g. Cress et al., 1996; Magliocchetti et al.,
1998, 1999). Furthermore, due to the low flux limits reached, these
surveys sample populations of sources, such as star-forming galaxies,
where the radio signal is produced by phenomena other than AGN
activity.

A complete investigation of the nature of radio sources and their
properties can only be achieved with multi-wavelength follow-up. In
particular optical identifications enable the acquisition of optical
spectra, which can be used to derive their spectral type and redshift
distribution (see e.g. Magliocchetti et al., 2000 - MA2000
hereafter). Great effort has recently been made to determine the
photometric and spectroscopic properties of radio sources at mJy
levels and fainter (Gruppioni et al., 1998; Georgakakis et al., 1999;
Sadler et al., 1999; MA2000; Masci et al., 2001). These studies have
proven extremely useful in characterizing the populations of radio
sources, but there are still significant uncertainties. For instance
it has been established that there is starbursting population of radio
sources, but there is still some debate about the overall fraction 
(see for instance the results from Sadler et al., 1999 as compared
with those derived in Gruppioni et al., 1998; Georgakakis et al.,
1999; MA2000).

These follow-up studies however suffer from two great limitations:
either they survey small areas of the sky and therefore generate
relatively small samples of objects, or they lack completeness. The
small number of sources leads to a low confidence level for any result
drawn from statistical studies, while radio or optical incompleteness
leads to biases in the determination of the redshift distribution of
radio sources at such low-flux levels.

In this Paper we present an analysis of the properties and angular
clustering of $\sim$ 4100 radio sources brighter than 1~mJy with
optical counterparts brighter $b_J \le 22$. This sample was obtained by
matching together objects from the FIRST and APM surveys over $\sim 350$
square degrees near the celestial equator.  Despite not having
redshift measurements for these sources, their joint radio,
photometric and morphological properties can be used to infer
extremely useful information on their nature.  Furthermore, for $b_J \le
21.5$, we obtain a catalogue of sources which is $\sim 100\%$ complete
in the optical band (MA2000) and 80\% complete in the radio band (with
a completeness level rising to 100\% for radio fluxes $S_{1.4 {\rm
GHz}}\ge 3$~mJy, Becker et al., 1995).

The photometric and morphological properties of the optical
identifications are then used to divide the whole sample of radio
sources into two well defined populations. The first population
consists of low-redshift radio galaxies and the second one of high-z
QSO and objects with no counterpart on the UKST plates.  The
homogeneity and completeness of the low-z sample allows us to study
the clustering properties of these sources.

Lastly, the area chosen for our analysis coincides with some of the
fields observed in the 2df Galaxy Redshift Survey (Maddox 1998,
Colless, 1999). This sample of optical identifications of $\sim$ 4100
radio sources therefore provides an excellent starting point for
further wide-area spectroscopic follow-up.

The layout of the paper is as follows: Section 2 briefly describes the
two surveys and the data coming from them.  Section 3 presents the
procedure we adopted to match radio and optical sources together,
while Section 4 is devoted to the analysis of the photometric
properties of the optical identifications. Section 5 examines the
clustering properties of the sample and Section 6 summarizes our
conclusions.  Throughout the paper we will assume $\Omega_0=0.4$,
$h_0=0.65$, $\Lambda=0.6$.

\section{The Catalogues}
\subsection{The FIRST survey}
The FIRST (Faint Images of the Radio Sky at Twenty centimetres) survey
(Becker, White and Helfand, 1995) began in the spring of 1993 and will
eventually cover some 10,000 square degrees of the sky in the north
Galactic cap and equatorial zones.  
The beam-size is 5.4~arcsec at 1.4~GHz, with an
rms sensitivity of typically 0.15~mJy/beam.  A map is produced for
each field and sources are detected using an elliptical Gaussian
fitting procedure (White et al., 1997); the $5\sigma$ source detection
limit is $\sim 1$~mJy. The astrometric reference frame of the maps is  
accurate to 0.05~arcsec, and individual sources have 90 per cent confidence 
error 
circles of radius $<$ 0.5~arcsec at the 3 mJy level, and 1~arcsec at the 
survey threshold. The surface density of objects in the catalogue
is $\sim 90$ per square degree, though this is reduced to $\sim 80$      
per square degree if we combine multi-component sources
(Magliocchetti et al., 1998). 
The depth, uniformity and angular extent of the survey are excellent
attributes for investigating, amongst others, the clustering properties 
of faint sources. The catalogue derived from this survey has been
estimated to be 95 per cent complete at 2~mJy and 80 per cent complete
at 1~mJy (Becker et al.,~1995).   
 
We used the 5 July 2000 version of the catalogue which contains
approximately 722,354 sources from the north and south Galactic caps.
This is derived from the 1993 through 2000 observations that cover
nearly 7988 square degrees of sky, and includes most of the area
$7^h20^m \simlt {\rm RA}(2000) \simlt 17^h20^m$, $22.2^\circ \simlt
{\rm Dec} \simlt 57.5^\circ$ and $21^h20^m \simlt {\rm RA}(2000)
\simlt 3^h20^m$, $-2.8^\circ \simlt {\rm Dec} \simlt 2.2^\circ$.

\subsection{The APM survey}
\begin{figure}
\vspace{8cm}  
\includegraphics{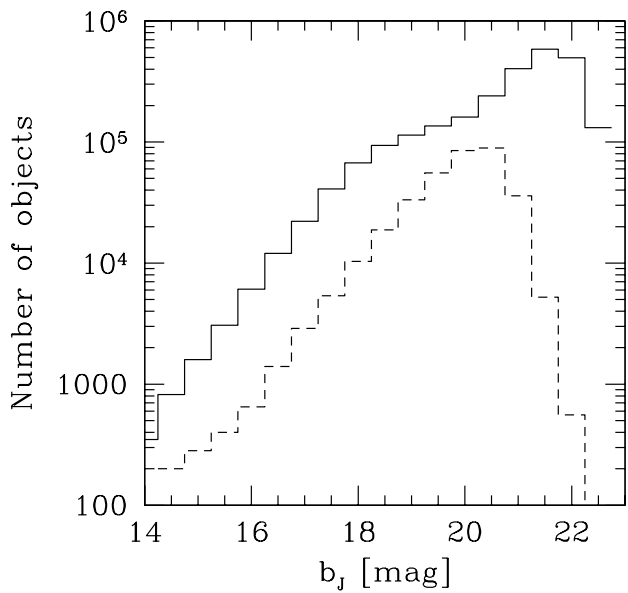}
\caption{Number of galaxies (dotted line) and stellar objects (solid
line) in the APM survey as a function of $b_J$ magnitude.
\label{fig:hist_all}}
\end{figure} 

The APM galaxy survey has been extensively described in Maddox et al.,
(1990a, 1990b, 1996). Briefly, it is based on APM scans of
$5.8^\circ\times 5.8^\circ$ UKST plates, covering about 7000 square
degrees in the regions Dec $<-20^\circ$ in the southern Galactic
cap and $-17.5^\circ\le$ Dec $\le 2.5^\circ$, $-7.5^\circ\le$ Dec $
\le 2.5^\circ$ respectively in the equatorial sgp and ngp.  A scan of
a typical plate records about 300,000 images, with a limiting
magnitude for image detection $b_J \simlt 22$. For each image, the
measurements show a positional accuracy of $\sim 1$~arcsec and an
isophotal magnitude accuracy of $\sim 0.1-0.2$~mag.

The photometry from each plate is corrected so that it is consistent
with the neighbouring plates, in the way described in Maddox et
al. (1990b).  The limit for uniform image detection over the whole
survey is set by the shallowest plate, and is found to be $b_J=21.5$.

\begin{table*}
\begin{center}
\caption{Number of sources on each UKST plate considered in this work;
the first column is for all the sources before any magnitude cut
applied to the APM catalogue (i.e. a non-uniform limit $21.5 \le b_J
\le 22$), while the second one illustrates the number of sources with
$b_J \le 21.5$ - the uniform completeness limit of the survey. Columns
3 and 4 are for the objects respectively identified as non-stellar and
stellar-objects (both for $b_J \le 20.5$).
\label{table_APM}}
\begin{tabular}{lllll}
Field Name& \# objs & \# objs with $b_J \le 21.5$ & \# galaxies ($b_J
\le 20.5$)  
& \# stars ($b_J \le 20.5$)\\
\hline    
f853  &        263346    &      115857  &         16096  &      78028\\
f854  &        207832    &      105631  &         15057  &      56021\\
f855  &        176135    &      108062  &         15715  &      54046\\
f856  &        159556    &      102762  &         14170  &      53507\\
f857  &        168511    &      103053  &         15031  &      53824\\
f858  &        133900    &      101223  &         14028  &      43720\\
f859  &        166885    &       98839  &         14201  &      46849\\
f860  &        228542    &       93941  &         11893  &      54350\\
f861  &        173320    &       96414  &         13003  &      41883\\
f862  &        176040    &      100065  &         12137  &      51545\\
f863  &        162137    &      119390  &         18089  &      53059\\
f864  &        233215    &      110147  &         13233  &      63483\\
f865  &        173488    &      123105  &         13544  &      65765\\
f866  &        227948    &      129765  &         14689  &      68062\\
f867  &        207689    &      145523  &         15225  &      75443   
\end{tabular}
\end{center}
\end{table*}   

Image profiles were then used to separate galaxies from stars; the
resulting sample of $\sim 3\cdot 10^6$ galaxies (APM Galaxy survey) is
$\sim 90-95$ per cent complete, and contamination from stellar objects
is about 5-10 per cent over the magnitude range $16\le b_J \le 20.5$.
For $b_J \le 20.5$ we then end up with two well defined subsamples
made by two different classes of sources, galaxies and stellar
objects.  Fainter than this, the measured galaxy profiles do not
greatly differ from stellar profiles and the very few images appear
significantly non-stellar. This resulting decrease in images
classified as galaxies can be seen in Figure \ref{fig:hist_all} where
we plotted the distribution of non-stellar (dotted line) and stellar
objects (solid line) as a function of the $b_J$ magnitude).
  
\begin{figure*}
\vspace{7cm}  
\includegraphics{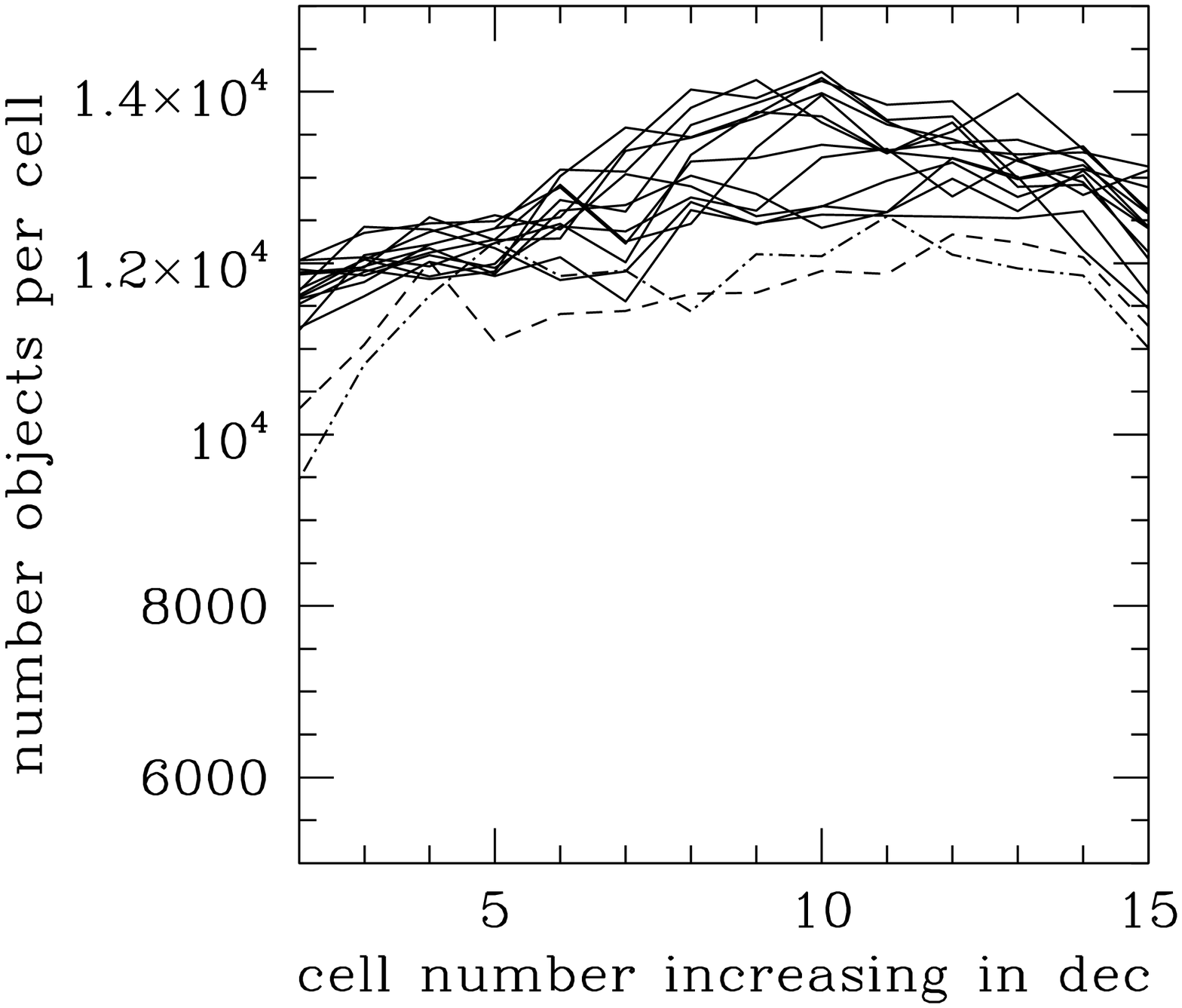}
\includegraphics{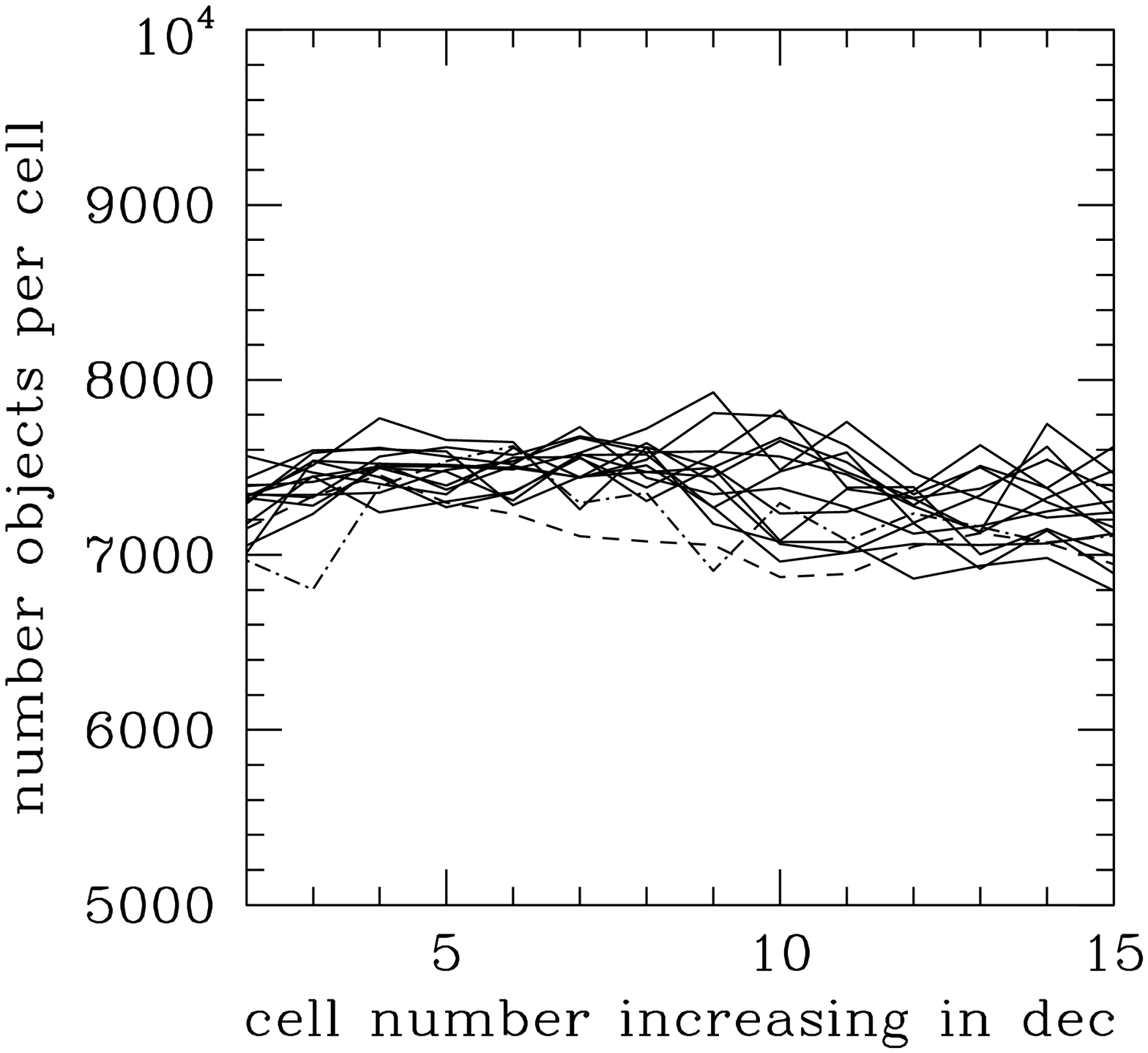}
\includegraphics{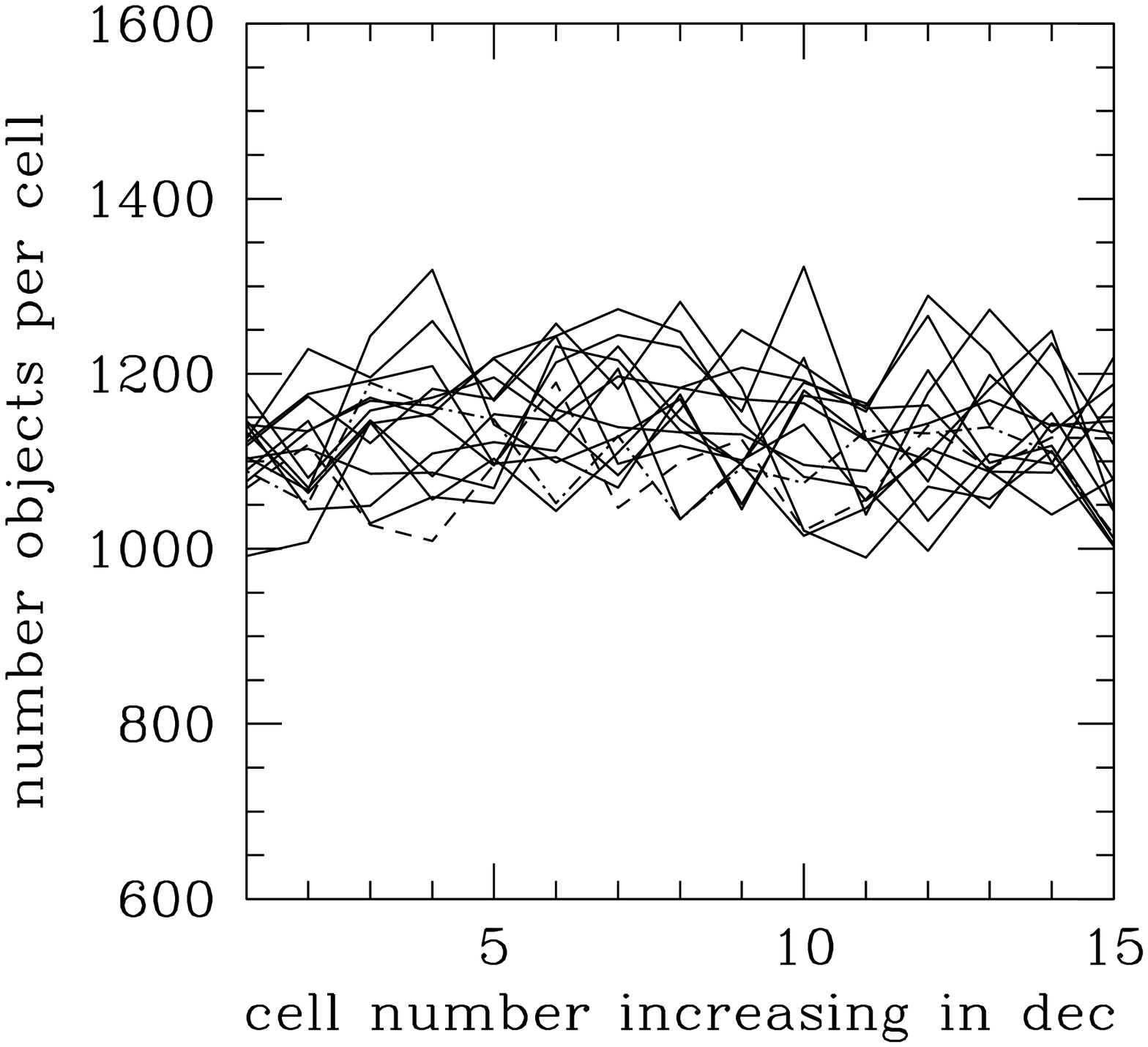}
\caption{Number of objects within each of the $15 \times 15$ sets of
cells $C_{ij}$, obtained by adding together sources from corresponding
cells on different UKST plates (see text for details), as a function
of cell number increasing with Dec. Each curve is derived by letting
the index $j$ vary between 1 and 15 while keeping the other index $i$
(indicating the RA) fixed.  Different curves are obtained for sets of
cells with different RAs.  Dashed and dashed-dotted lines correspond
to the sets of cells respectively placed on the left-hand and
right-hand side of each plate.  The panel on the left-hand side is for
sources with $b_J \le 22$, the one in the middle for $b_J \le 21.5$,
while the one on the right-hand side is for objects identified as
galaxies with $b_J \le 20.5$.
\label{fig:cells}}
\end{figure*}  

\subsection{Uniformity of the APM Data}

In order to test for systematic effects across the plates we have
analysed the surface density of images as a function of position
within each plate.  We divided each plate into a grid of $15\times 15$
$c_{ij}$ cells in RA and Dec and counted the number of objects in each
cell, summing corresponding cells on different plates together.  In
the absence of systematic effects, the number of sources in each of
the cells $C_{ij}=\sum_{\rm Plates}c_{ij}$ should be the same within
Poisson errors, since superposing sources from different regions of
the sky will average over any real structure in the distribution.

The resulting cell counts are presented in Figure \ref{fig:cells},
where each line shows the cell count plotted against Dec index $j$ for
fixed RA index $i$, and the different lines represent sets of cells
with different RAs within each plate.  The left-hand panel shows the
number of objects within each set of cells $C_{ij}$ for all the
sources detected i.e. with a non-uniform detection limit, varying
between $21.5 \simlt b_J \simlt  22$.


There are clearly systematic differences in the average number of
images in different parts of a plate.  In particular all the lines are
lower at the ends that at the centres, showing that there are fewer
objects at the edges of the each plate than in the centre. This is
because of the geometrical vignetting and differential desensitization
of the photographic emulsion which increase the image detection limit
near the plate edges. The effective transmission is quite flat close
to the telescope axis but falls to $\simlt 70$ per cent of the central
value at the edges of the plates. The dashed and dot-dashed lines
correspond to the sets of cells respectively placed on the outermost
regions on the left-hand and right-hand side of each plate and are
$\sim 30\%$ lower than the central values. A correction has been
applied to make the measured magnitudes uniform over the field, but
this was applied after the image detection process, and so we do not
recover any images lost below the original detection limit at the
edges of each plate.  However, if we adopt a magnitude cut of
$b_J=21.5$, this is brighter than the original detection limit even in
the plate corners, and so we obtain a uniformly selected sample of
images.  The middle panel in Figure \ref{fig:cells} shows the number
of sources with $b_J\le 21.5$, and it is clear that the a catalogue of
sources is uniform across the sky.  The subsample of objects
identified as galaxies also appears uniform, as shown in the
right-hand panel of Figure \ref{fig:cells}.

\subsection{Optical colours} 

The Maddox et al (1990) version of the APM data includes magnitude
measurements only in the $b_J$ band, but Irwin et al. have subsequently
scanned UKST $R$ plates for the same area of sky. This more recent data
can be found in the APMCAT database at {\tt
http://www.ast.cam.ac.uk/$^\sim$apmcat} and includes both $b_J$ and $R$
magnitudes for objects down to $b_J \sim 22$, $R \sim
21$. Unfortunately for us, the APMCAT data was processed with stellar
photometry in mind, and this leads to magnitude estimates that are
significantly different to the galaxy magnitudes of Maddox et
al. Re-processing all of the $R$ data to make it consistent with the
Maddox et al $b_J$ data would be a major undertaking, so we used a
simple approximation to estimate the galaxy $R$ band magnitudes. We make
the assumption that $b_J-R=b_{J_{\rm APMCAT}} -R_{_{\rm APMCAT}}$ which
assumes that the difference between the stellar and galaxy magnitudes
is the same in $b_j$ as in $R$ for all plates. This is not exactly
true but is probably not in error by more than 0.1 magnitudes. Thus we
estimate the galaxy $R$ magnitudes as $R=b_J-[b_{J_{\rm APMCAT}}-R_{_{\rm
APMCAT}}]$.

\section{Matching procedure}

The FIRST and APM surveys only overlap in a relatively small region of the 
sky set on the equatorial plane between 9$^h$ 48$^m$ $\simlt$ RA(2000) 
$\simlt$ 14$^h$ 32$^m$ and -2.77$^\circ$ $\simlt$ Dec $\simlt$ 2.25$^\circ$.
This corresponds to the UKST survey fields f853 to f867. 
The number of optical sources observed within this area down to the magnitude 
limit of the APM survey is $\sim 2.8 \cdot 10^6$; the number of radio objects 
included in the FIRST survey down to 1~mJy within the same area is $\sim 
24,000$. Table \ref{table_APM} shows the number of sources in the APM 
catalogue appearing in each field; first and second column are respectively 
obtained for $b_J \le 22$, magnitude limit of the survey, and $b_J \le 21.5$, 
completeness limit of the survey. Columns 3 and 4 give the numbers of 
galaxies and stellar objects - obtained as in Maddox et al. (1990a) for 
$b_J \le 20.5$ - on each of the plates considered in this work.

We identified the optical counterparts of FIRST radio sources by
matching together sources from the radio sample with those from the
optical catalogue and simply taking pairs with separation less than a
chosen radius as the correct identifications. The maximum acceptable
radius depends on the accuracy of the measured positions.  As
mentioned in Section 2, the positional accuracy in the FIRST survey is
estimated to be $\simlt$0.5 arcsecs at the 3 mJy level, reaching the value of 
1 arcsec only at the
survey threshold (Becker, Helfand \& White, 1995). This makes a mean 
positional error across the survey of about $\sim 0.5$~arcsec; 0.5 
arcsec is also  
the position accuracy for objects in the UKST plates down to $b_J
\simeq 21.5$ (Maddox et al., 1990a), so that the expected value for
the radius to be used in the matching procedure is $\sim \sqrt{2\times 0.5^2}
\simeq 0.7$~arcsec. However, radio and optical reference frames may also be
offset with respect to each other. For example MA2000 find, for a
subsample of sources drawn from the FIRST survey, a mean positional
offset of about 0.8 arcsec between optical (APM) and radio reference
frames. Lastly it should also be noted that the centre of radio
emission is often displaced from the centre of optical emission,
especially when the radio source is extended. These two further
effects generate an extra error term which has to be added in
quadrature to the positional accuracies in both radio and optical
catalogues, giving rise to a matching radius of the order of $\sim
0.8-0.9$ arcsecs.

Tackling the problem from a more pragmatic point of view, we
considered the distribution of the residuals $\Delta x=x_{\rm
RADIO}-x_{\rm OPTICAL}$, $\Delta y=y_{\rm RADIO}-y_{\rm OPTICAL}$
between the positions of all radio and optical pairs with separations
$|\Delta x|$ and $|\Delta y|$ less than 5 arcsecs.

\begin{figure}
\vspace{8cm}  
\includegraphics{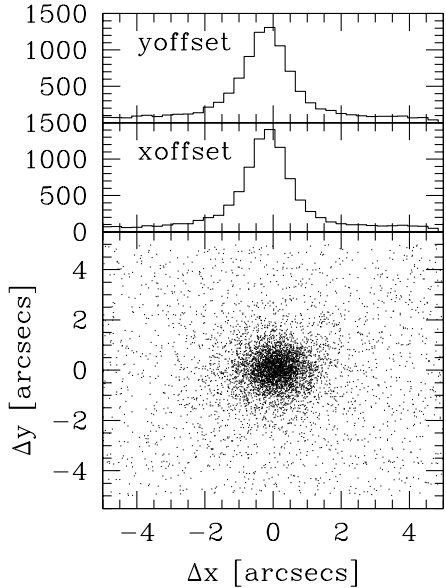}
\caption{Lower panel: distribution of the residuals $\Delta 
x=x_{\rm RADIO}-x_{\rm OPTICAL}$, 
$\Delta y=y_{\rm RADIO}-y_{\rm OPTICAL}$ between radio and optical positions.
Middle and top panels: number of matches as a function of $\Delta x$ and 
$\Delta y$ offsets.   
\label{fig:allres1}}
\end{figure}

The results are presented in Figure \ref{fig:allres1}, where the
bottom panel shows the $\Delta x$ - $\Delta y$ distribution with a
point for each radio-optical pair, while the middle and top panels
respectively show histograms of number of matches as a function of
$\Delta x$ and $\Delta y$ offsets.
The true optical identifications show up as a highly peaked
concentration of points near zero offset, with a rms value of the
distributions (both in $\Delta x$ and $\Delta y$) of about 0.7
arcsec. Note that this value is
very close to the one formerly obtained by following a simple
qualitative analysis, and encloses $\sim 70$ per cent of the candidate
matches found within 5 arcsecs. It follows that a 2 arcsec match radius 
is equivalent to about $2.5\sigma$, and should then include $\sim 97\% $ of 
the true identifications. 
We therefore assume all the objects with positional residuals
greater than 2 arcsecs to be random coincidences, and choose the value
of 2 arcsecs as the correct matching radius. 
\begin{table*}
\begin{center}
\caption{Number of radio sources on each UKST plate considered in this work; 
the first column 
shows the total number of radio sources  from the FIRST survey, 
while the second and third ones illustrate the number of sources with an 
optical identifications on the plates, respectively for $b_J \simlt 22$ - 
magnitude limit of the survey - and $b_J \le 21.5$ - completeness limit of the 
survey. Columns 4 and 5 are for radio objects respectively 
identified as galaxies and stellar images (both for $b_J \le 20.5$) 
\label{table_identifications}}
\begin{tabular}{llllll}
Field Name& \# radio & \# identifications  & \# identifications ($b_J \le 21.5$)&
\# galaxies ($b_J \le 20.5$) & \# stellar ($b_J \le 20.5$)\\
\hline                                   
f853  &1584 &    304&        215&           105&               47\\
f854  &1602 &    279&        214&           104&               44\\
f855  &1531 &    234&        183&           105&               29\\
f856  &1578 &    252&        208&            89&               53\\
f857  &1589 &    255&        214&          106&                45\\
f858  &1579 &    234&        204&            90&               41\\
f859  &1690 &    295&        234&           123&               39\\
f860  &1588 &    278&        196&            80&               55\\
f861  &1576 &    275&        199&            94&               36\\
f862  &1582 &    268&        191&            83&               49\\
f863  &1695 &    258&        225&           130&               42\\
f864  &1730 &    287&        193&            78&               45\\
f865  &1613 &    272&        223&            98&               64\\
f866  &1788 &    307&        249&           115&               44\\
f867  &1674 &    277&        228&            98&               45\\          
\end{tabular}
\end{center}
\end{table*}       

Columns 2 and 3 of Table \ref{table_identifications} respectively give
the total number of radio sources and number of sources with optical
identifications (with $b_J \simlt 22$) found in each field. The total
number of sources in the area $148.13^\circ \simlt \rm{RA (2000)}
\simlt 218.13^\circ$, $ -2.77^\circ \simlt \rm{Dec} \simlt 2.25^\circ
$ with an identified optical counterpart within 2 arcsecs is 4075,
16.7 per cent of the whole radio sample. This percentage drops to 13
per cent (corresponding to 3176 identifications, column 4 of Table
\ref{table_identifications}) if we include the further constraint $b_J
\le 21.5$, to obtain a uniform sample from the APM survey.

\begin{figure}
\vspace{8cm}  
\includegraphics{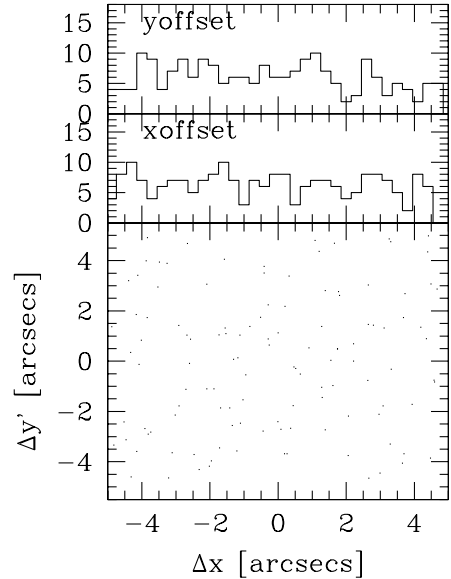}
\caption{Lower panel: distribution of the residuals $\Delta 
x=x_{\rm RADIO}-x_{\rm OPTICAL}$, 
$\Delta y'=y'_{\rm RADIO}-y_{\rm OPTICAL}$ between radio and 
optical positions, 
where a vertical shift of 1 arcmin has been added to all the radio positions.
Middle and top panels: number of matches as a function of $\Delta x$ and 
$\Delta y'$ offsets.   
\label{fig:allres2}}
\end{figure}      

In order to quantify how many matches are likely to be random
coincidences, we ran the matching code once again, after shifting all
the radio positions 1 arcmin north. The results are shown in Figure
\ref{fig:allres2}. The positional residuals are uniformly distributed
on the $\Delta x$ $\Delta y'$ plane as expected for random
coincidences.  Furthermore, we find only 225 sources with offsets
within our chosen matching radius of 2 arcsecs, corresponding to $\sim
5$ per cent of the true matches.  Contamination from spurious
coincidences at this level is not a problem in our analysis.  Note
that if we had chosen a matching radius as large as 5 arcsecs, the
percentage of random matches would have been 1200/5866 $\simeq 20$ per
cent, which would have required a more sophisticated treatment of
contamination. These numbers are in good agreement with 
the results obtained by evaluating the expected number of chance coincidences: 
for 68.7 radio sources per square degree, one in fact respectively obtains 
182 and 1139 random matches by considering radii of 2 and 5 arcsecs over the 
350 square degrees of analysed area.

\begin{figure}
\vspace{8cm}  
\includegraphics{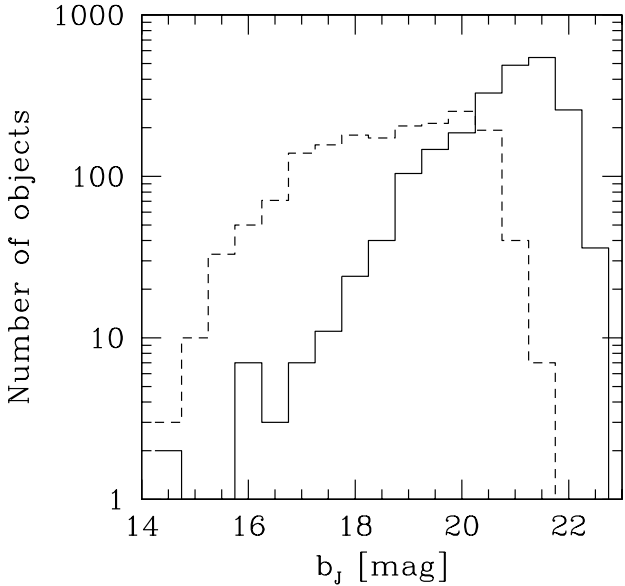}
\caption{Number  
 of galaxies (dotted line) and stellar-like objects (solid line) in the radio 
subsample with optical identifications as a function of $b_J$ magnitude.
\label{fig:hist_identifications}}
\end{figure} 

\section{Optical Properties of Radio Sources}

We divided the radio sources with optical identifications into
galaxies and stellar objects based on the optical classifications
described in Section 2. Figure \ref{fig:hist_identifications} shows
the number of galaxies (dotted line) and stellar sources (solid line)
in the radio subsample with optical identifications (hereafter called
the id-sample) as a function of $b_J$ magnitude.  As expected from the
analysis performed in Section 2, the number of sources identified as
galaxies rises with increasing magnitude up to $b_J \sim 20.5$, where
the optical images become too faint to allow the detection of any
extended structure. Correspondingly, the number of stellar objects
increases beyond its real value for $b_J \simgt 20.5$. Again $b_J
=20.5$ is seen as the faintest magnitude for which we can discriminate
between extended images (radio galaxies) and stellar images (mainly
QSO). The stellar identifications also include a very small number of
galactic stars - less than 0.1 per cent of the objects in the FIRST
survey are expected to be associated with stars, see Helfand et al.,
1997). Columns 5 and 6 of Table \ref{table_identifications} indicate
the number of non-stellar and stellar sources found on each of the
UKST plates in our analysis.

\begin{figure}
\vspace{8cm} 
\includegraphics{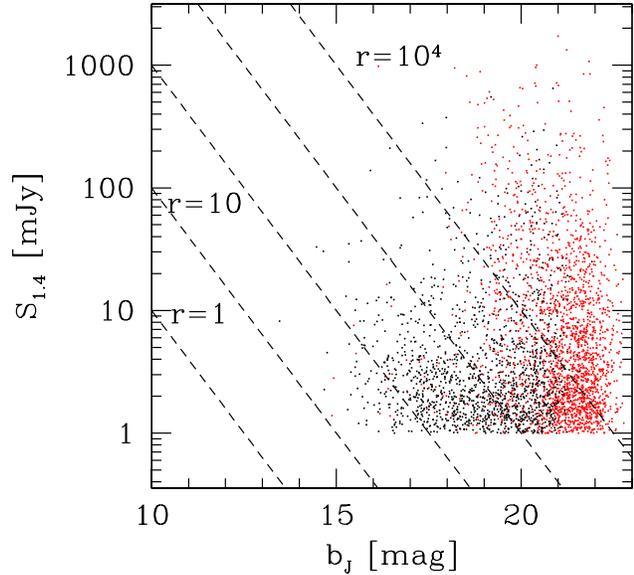}
\caption{Radio flux S vs. $b_J$ magnitude. The dashed lines are the loci of 
constant radio-to-optical ratio r$_B$ (see text for details). Red points 
indicate stellar-like sources (i.e. QSO), while the black ones are for objects 
identified as galaxies. 
\label{fig:FB}}
\end{figure}

\begin{figure}
\vspace{8cm}  
\includegraphics{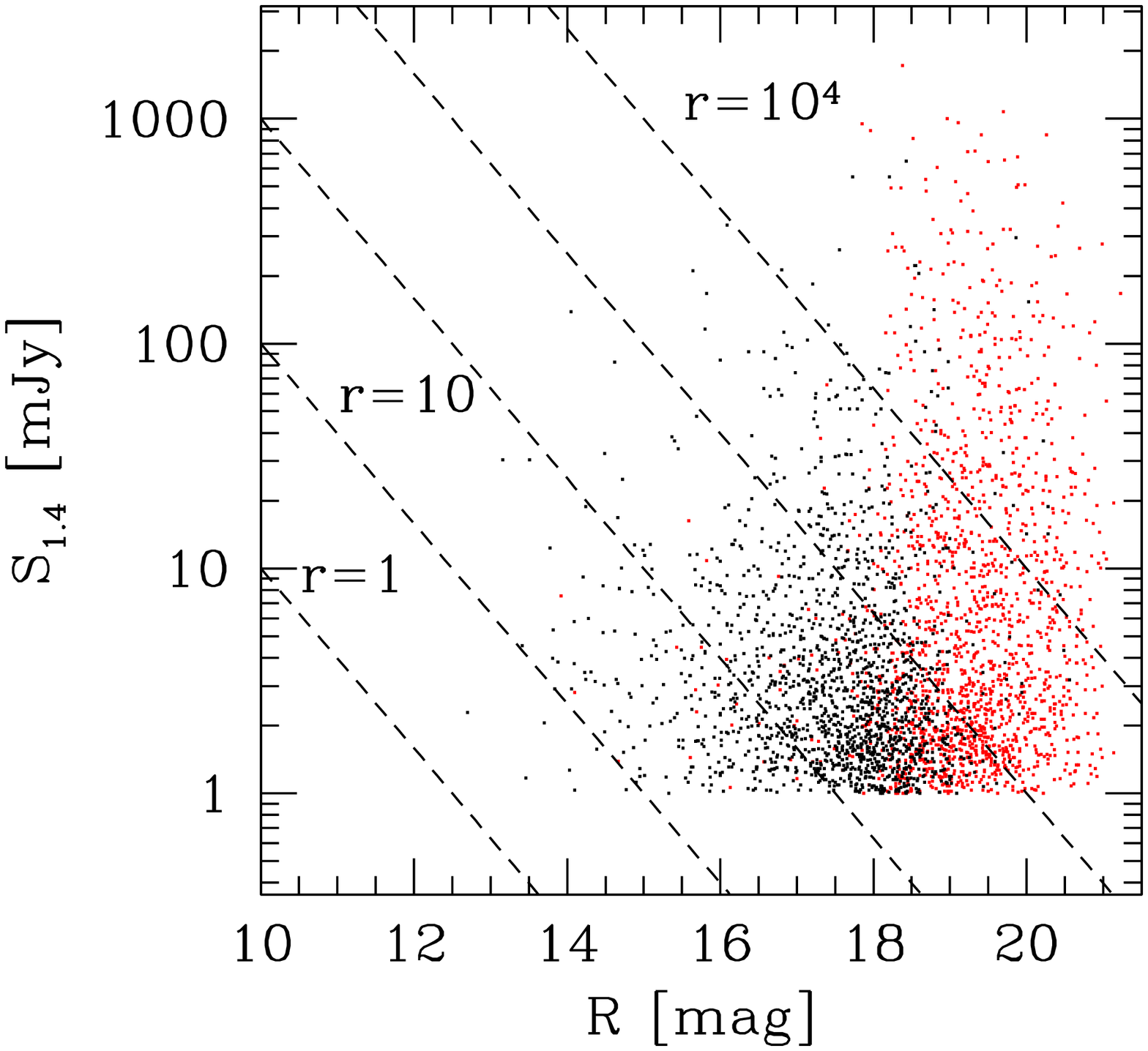}
\caption{Radio flux S vs. R magnitude. The dashed lines are the loci of 
constant radio-to-optical ratio r$_R$ (see text for details). Red points 
indicate stellar-like sources (i.e. QSO), while the black ones are for objects 
identified as galaxies. 
\label{fig:FR}}
\end{figure}

A joint analysis of the radio and photometric properties of the
id-sample can provide us with extremely useful insights on the nature
of the radio sources under examination.  The $b_J$ and $R$ magnitudes
are plotted against radio fluxes in Figures \ref{fig:FB},
\ref{fig:FR}, where black dots indicate galaxies and red dots indicate
stellar objects. Note that beyond $b_J \sim 20.5$ (corresponding to $R$
$\sim 19.5$) the distinction between galaxies and stars becomes very
difficult, therefore all the sources appear in the catalogue as
stellar. The following analysis will therefore only apply to
magnitudes $b_J \le 20.5$. 

\begin{figure}
\vspace{8cm}  
\includegraphics{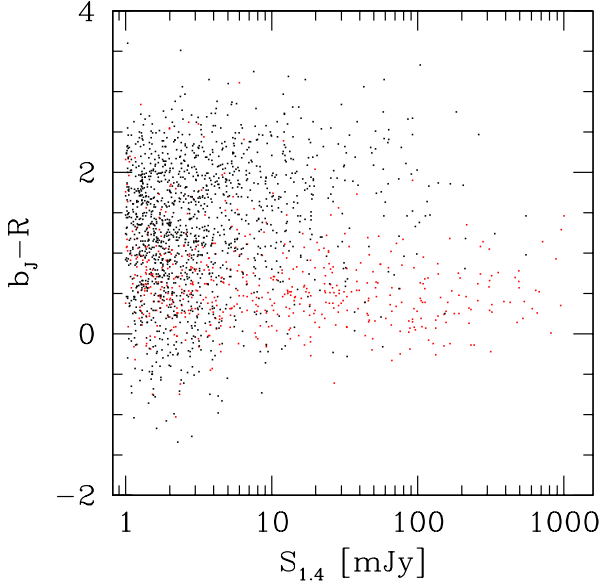}
\caption{Colour $b_J-R$ vs flux for all the sources in the id-sample 
with $b_J \le 20.5$. Red points 
indicate stellar-like sources (i.e. QSO), while the black ones are for objects 
identified as galaxies. 
\label{fig:FRB}}
\end{figure}  

\begin{figure}
\vspace{8cm}  
\includegraphics{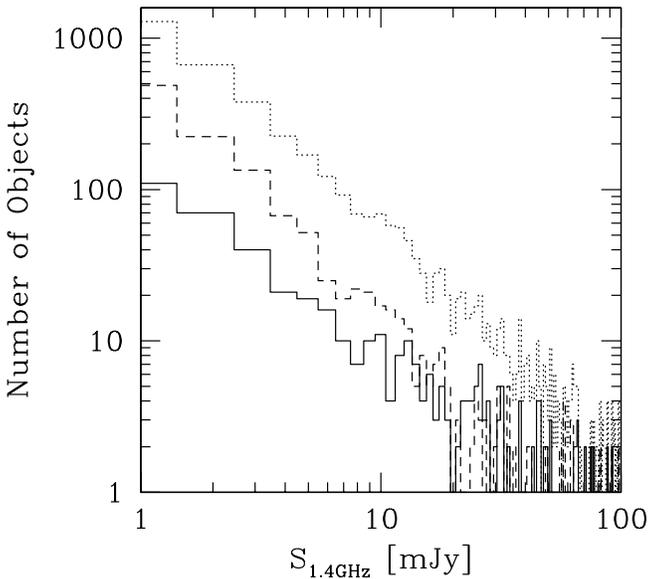}
\caption{Number of sources per flux interval in the id-sample. 
The dotted line represents the whole sample, while the dashed and solid lines 
respectively are for galaxies and stellar-like sources (with $b_J \le 20.5$)
\label{fig:hist_flux}}
\end{figure}  

The radio-to-optical ratio is defined as r$_{\rm m}$ $=\rm{S}\times
10^{({\rm m}-12.5)/2.5}$, where S is the radio flux in mJy and m is
the apparent magnitude (either $R$ or $b_J$ depending on the specific
case. We recall here that a source is considered radio-loud if r$_B$
$\simgt 10$ (see e.g. Urry \& Padovani 1995). It has been shown (see
for instance Magliocchetti, Celotti \& Danese, 2001) that for
star-bursting galaxies, where the star-formation activity is so
intense that it hides the signatures of any central BH, r$_B$ $\simlt
100$. Thus the small number of sources below the r$_B$=100 line in
Figure \ref{fig:FB} shows that there are $<100$ star-forming galaxies
in the FIRST sample down to 1 mJy. This is further confirmation of the
result that star-bursting sources show up only at radio fluxes below
the mJy level (see for instance, Gruppioni et al. 1998, Georgakakis et
al. 1999 and MA2000).

The region of the S-$b_J$ plane between r $\simeq 100$ and r $\simeq
10^4$ is mainly dominated by sources identified as (radio) galaxies,
and this population constitutes the majority of the id-sample. For r
$\simgt 10^4$ and $b_J\le 20.5$ stellar identifications, most likely
to be high-z QSO, dominate the sample. Fainter than $b_J=20.5$ this
effect is mainly an artefact of the data because the optical images 
are too faint to resolve galaxies.

Figure \ref{fig:FR} illustrates the flux-magnitude diagram in the $R$
band.  Apart from the features already seen in Figure \ref{fig:FB}, we
note an interesting shift of the population of radio galaxies towards
brighter red magnitudes. This shift is especially significant for S
$\simgt 10$ mJy and is not observed for the stellar identifications,
highlighting the fact that radio-galaxies are typically red,
early-type objects.

This effect can be seen more clearly in Figure \ref{fig:FRB}, which
plots the $b_J-R$ colour of sources in the id-sample for objects with
$b_J$ $\le 20.5$. The majority of the id-sample consists of galaxies
with very red colours, up to $b_J -R \sim 4$. This conclusion is
actually made even stronger as in our analysis we might have missed
very red sources which had $b_J$ magnitudes too faint to be included
in the original catalogue. The id-sample was drawn from a catalogue
selected on $b_J$ magnitude; the red magnitudes were added later as
explained in Section 2.  Note the clear distinction in Figure
\ref{fig:FRB} between radio galaxies, which appear mostly with $b_J-R
\simgt 1$ and QSO, which dominate the region $0 \simlt b_J-R
\simlt 1$.  This division becomes striking for radio fluxes S
$\simgt$ 10, where the two distributions take the shape of a tuning
fork.  Again we can see that star-forming galaxies showing blue
(i.e. for $b_J-R \simlt 0$) colours and low (S $\simlt 10$)
radio-fluxes constitute only a few tens of objects in the lower part
of the plot.

Also when we consider the number of sources per unit of radio flux
(see Figure \ref{fig:hist_flux}) we see that stellar and non-stellar
ids show different behaviours. The number of galaxies (the dashed
line), shows a steeper slope than the number of QSO (the solid
line). This difference reflects the fact that galaxies tend to have
lower radio fluxes, while QSO dominate the region $S_{1.4 {\rm
GHz}}\simgt 20$~mJy.

\section{Spatial Distribution and Clustering Properties of the Sample}

\begin{figure*}
\vspace{10cm}  
\includegraphics{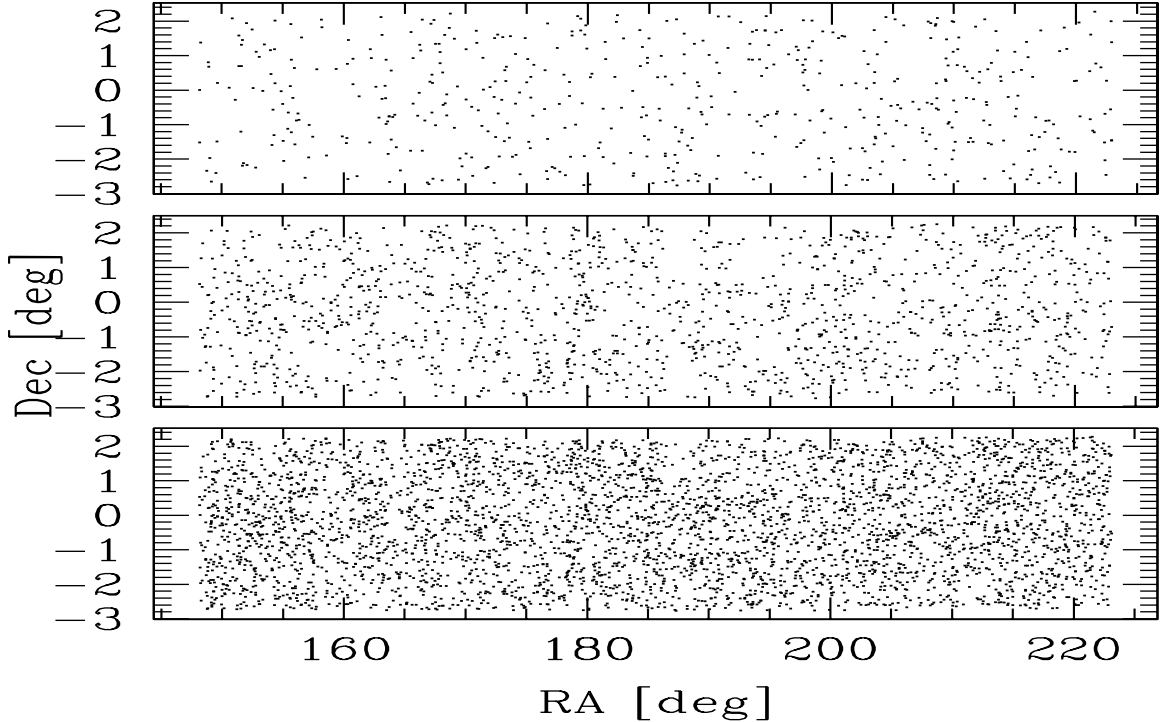}
\caption{Projected distribution of radio sources with optical identifications: 
bottom panel for the whole sample ($b_J \le 21.5$), middle panel for the 
sources identified as 
galaxies ($b_J \le 20.5$) and top panel for those identified as stellar 
($b_J \le 20.5$). 
\label{fig:distrib}}
\end{figure*} 

The acquisition of optical identifications for radio sources is not
only useful in providing optical information, it also gives the
opportunity to make a distinction between low-redshift sources and
more distant ones.

Passive radio galaxies are known to be reliable standard candles (see
e.g. Hine \& Longair, 1979; Rixon, Wall \& Benn, 1991). The mean
absolute magnitude in the red band is $M_R\simeq -23$ (see e.g. MA2000), and
there is little scatter about the mean (MA2000 estimate $\Delta$M$_R 
\simeq 0.25$ at the 1$\sigma$ level). Thus measurements of the apparent 
magnitude $R$ for these
sources can provide us with an estimate of their redshifts according
to the relation
\begin{eqnarray}
R-M_R=-5+5{\rm log}_{10}d_L (pc)
\label{eq:R}
\end{eqnarray}       
where $d_L$ the luminosity distance.  
The id-sample is complete to $R \sim 19.5$, and from this formula we
expect that objects brighter than $R \sim 19.5$ should be found in the
redshift range $0\simlt z\simlt 0.3$.\\ As we saw in the previous
Section, the optically extended id-sample is mostly made up of
early-type galaxies, with star-bursting objects contributing only a
small fraction. So we can apply equation (\ref{eq:R}) to the whole
subsample of radio galaxies, and deduce that it contains radio-sources
that are mostly closer than $z\simeq 0.3$.  The spread in the $R-z$
relation means that the upper redshift limit will be a gradual cutoff;
see also MA2000.

The other sources that appear in the id-sample are mainly QSO with a
very small contamination from nearby stars.  Unfortunately, we have no
simple way to distinguish between these two classes of objects, or to
estimate the QSO redshift. However, apart from rare exceptions, we
expect QSO to be all placed at high redshifts, at least beyond z
$\simgt 0.3$.\\ The optical distinction between point-like and
extended sources as discussed in the previous Section, allows us to
divide the id-sample in a low-z subsample - comprising mostly of
early-type galaxies and roughly complete to $z\simeq 0.3$ - and a
sample of QSO mostly found at redshifts well beyond 0.3.

Figure \ref{fig:distrib} shows the angular distribution of these
sources projected on the sky. The bottom panel represents the
distribution of all the objects in the id-sample with $b_J \le 21.5$,
limit for completeness and uniformity of the optical catalogue, while
the middle one includes radio sources identified as galaxies and the
top one is for point-like sources (both in the case of $b_J \le
20.5$).

We can quantify the level of clustering in each sample by means of the
two-point correlation function. Ideally one would like to obtain the
spatial correlation function $\xi(r)$ but, since we do not have
measured redshifts for any sources in the id-sample, we have to deal
with its angular counterpart $w(\theta)$. We recall here that the
angular two-point correlation function measures the excess
probability, with respect to a random Poisson distribution, of finding
two sources in the solid angles $\delta\Omega_1$ $\delta\Omega_2$
separated by an angle $\theta$.  It is defined as
\begin{eqnarray}
\delta P=n^2\delta\Omega_1\delta\Omega_2\left[1+w(\theta)\right]
\label{eqn:wthetadef}
\end{eqnarray}
where $n$ is the mean number density of objects in the catalogue under
consideration.        

We calculated $w(\theta)$ using the estimator (Hamilton, 1993)
\begin{eqnarray}
w = &\frac{4DD \ RR}{DR^2}  &-1,
\label{eqn:wtheta_ests}       
\end{eqnarray} 
where the number of data-data, random-random and data-random pairs
separated by an angle $\theta$ are denoted by DD, RR and DR
respectively.\\ Random catalogues were generated with a spatial
distribution modulated by both the APM and FIRST coverage maps, so
that the instrumental window functions do not affect the measured
clustering.  We estimate the errors on the $w(\theta)$ measurements by
assuming the the pair counts follow Poisson statistics. This
underestimates the uncertainties, but we do not expect this bias to be
a large factor.  However, the errors in $w(\theta)$ at each point are
correlated, so it is easy to overestimate the significance of any
features apparent in $w$.

\begin{figure}
\vspace{8cm}  
\includegraphics{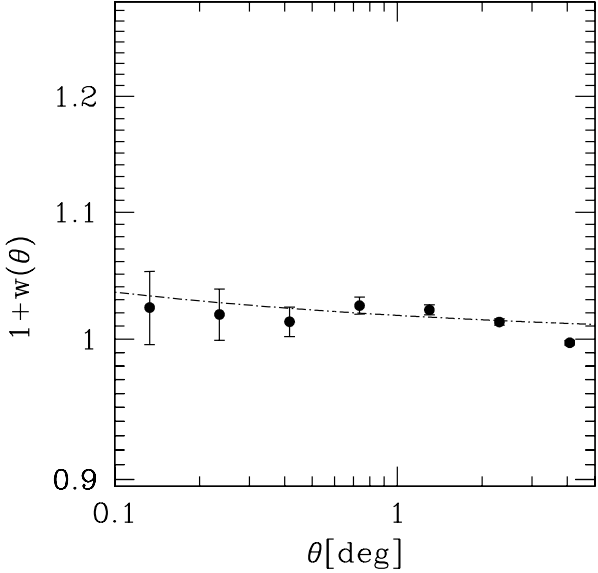}
\caption{The angular correlation function $1+w(\theta)$ for the whole sample 
of radio sources with an optical identification within $b_J=21.5$. 
Error-bars are obtained as Poissonian estimates for the catalogue under 
consideration. The dashed-dotted 
line shows the best fit to the data and is obtained for a functional form 
$w(\theta)=A\: \theta^{1-\gamma}$, where $A\sim 0.018$ and $\gamma\sim1.3$. 
\label{fig:wall}}
\end{figure} 

In Figure \ref{fig:wall} we show the results for $w(\theta)$ of the
whole id-sample with $b_J \le 21.5$. As already stated, error-bars are given 
by Poisson estimates for the catalogue under consideration.  The clustering
signal is significantly greater than zero at all angular scales;  
if we assume a power-law form for $w(\theta)$, $w(\theta)=A\:
\theta^{1-\gamma}$, we find, via a $\chi^2$ fit to the data, $A\sim
0.018$ and $\gamma\sim1.3$.\\
However, as we have already seen, the id-sample is a hybrid mixture of
low- and high-z sources, which means that this measurement is not
particularly relevant. More interesting information can be derived
from the analysis of the clustering signal produced by the two
different classes of objects, namely stellar and extended sources.

\begin{figure}
\vspace{8cm}  
\includegraphics{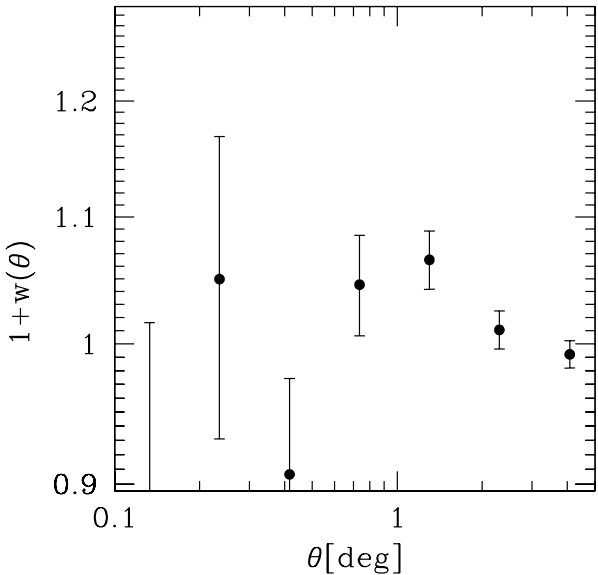}
\caption{The angular correlation function 1+$w(\theta)$ for the subsample 
of radio sources (with $b_J\le 20.5$) showing an optical compact structure. 
Error-bars are as in Figure \ref{fig:wall}.
\label{fig:wstars}}
\end{figure} 

We then estimated $w(\theta)$ for the stellar-like sources (535
objects with $b_J \le 20.5$), and the resulting $1+w(\theta)$ is shown
in Figure \ref{fig:wstars}. The clustering signal is dominated by
large errors due to the small number of objects. Nevertheless the
signal is positive at almost all angular scales up to $\theta\sim
5^\circ$. 

\begin{figure}
\vspace{8cm}  
\includegraphics{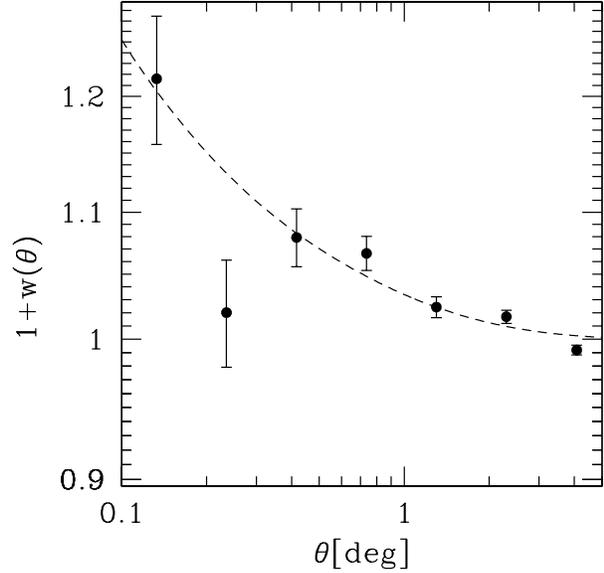}
\caption{The angular correlation function 1+$w(\theta)$ for the subsample 
of radio galaxies (with $b_J \le 20.5$). Error-bars are 
as in Figure \ref{fig:wall}. The dashed line represents 
predictions from clustering evolution models (see text for details).
\label{fig:wgals}}
\end{figure}

The most interesting result is represented by the angular correlation 
function for the 1494 radio galaxies in the id-sample. As illustrated by 
figure \ref{fig:wgals}, in this case the signal is quite 
strong, showing once again that radio sources are more clustered than
normal galaxies (see also Peacock \& Nicholson, 1991; Cress et al.,
1996; Loan, Wall \& Lahav, 1997; Magliocchetti et al., 1998:
Magliocchetti et al., 1999). In fact, if we parameterize $w(\theta)$ as a 
power-law: $w(\theta)=A\:\theta^{1-\gamma}$, we find - via a $\chi^2$ fit to 
the data - $A\sim 0.03$ and $\gamma\sim 2.1$, with a value for the slope close 
to those obtained for samples of early-type galaxies only, known to be more 
strongly clustered than other galaxy populations (see e.g. Maddox et 
al., 1990c and Loveday et al., 1995). Note that this result also agrees with 
the observational evidence for AGN-fuelled radio sources to be hosted by 
elliptical/early-type galaxies (see e.g. McLure et al., 1999).\\ 

Since this last measurement has been obtained for a homogeneous
sample of sources with a fairly well determined redshift distribution,
we can predict the clustering for a range of models to be directly
compared with the data. The first step in our model is to calculate
the mass-mass correlation function $\xi_m(r,z)$. We use the analytical
form derived using the Peacock \& Dodds (1996) approach, extended to z
$\ne 0$ by Matarrese et al. (1997). This starts with a primordial CDM
power spectrum (COBE normalized in our case; Bunn \& White, 1997) and
writes the non-linear part of the power-spectrum $P_{NL}(k_{NL})$ as a
function of its linear part $P_L(k_L)$, where $k_L$ and $k_{NL}$ are
the linear and non-linear wave-number respectively (see Magliocchetti
et al. (2000a) for a complete derivation of the various quantities).

The predicted angular correlation function $w_{th}(\theta)$ can then
be obtained from $\xi_m(r,z)$ using the relativistic Limber equation
(Peebles, 1980) \begin{eqnarray}
w_{th}(\theta)=2\;\frac{\int_0^{z_{max}}N(z)^2\; \frac{dx}{dz}\;
b(z)^2\;dz
\int_0^{\infty}\xi_m(r,z)\;du}{\left[\int_0^{z_{max}}N(z)\;dz\right]^2},
\label{eq:limber}
\end {eqnarray}
where $x$ is the comoving coordinate, $u$ is related to the spatial
distance $r$ between two sources separated by an angle $\theta$ (in
the small angle approximation) via $r\simeq
\left(u^2/F^2+x^2\theta^2\right)^{1/2}$ ($F$ is the curvature
correction), $N(z)$ is the redshift distribution for the sources under
consideration, $z_{max}$ is the maximum redshift of the sample and
$b(z)$ is the bias function which determines the way radio sources
trace the mass distribution. Note that, by imposing a $z_{max}$ we have assumed
that the sample completeness is a step function, which immediately
drops to zero beyond the redshift where the sample starts to become
incomplete (0.3 in our case).  We use the $N(z)$ obtained by MA2000
(Fig. 13). This is based on redshift measurements for 76 radio sources
selected from both the FIRST survey and the Phoenix survey (Hopkins et
al., 1998; Georgakakis et al., 1999) surveys. These samples were
selected at a limit of 1 mJy and include redshift measurements up to z
$\simeq 0.5$.

A little more attention has to be devoted to the issue of bias and its
evolution. Of the various models presented in literature, the
so-called {\it merging model} (Mo \& White, 1996; Matarrese et al.,
1997; Moscardini et al., 1998; Martini \& Weinberg, 2001) has the most
appealing physical motivation: dark matter halos simply undergo
dissipative collapse (i.e. merging).  However this model has been
shown to not fit the clustering data at low-redshifts (see e.g. Baugh
et al., 1999; Magliocchetti et al., 1999; Magliocchetti et al., 2000a
to mention just a few). The reason for the discrepancy may be that the
merging rate of objects is very high for z $\simgt 1$ but becomes
negligible for z $\simlt 0.5$.\\ A better bias model to describe the
data in this situation is the so-called {\it galaxy conserving model}
(Fry, 1996). This model assumes that galaxies form at some epoch and
then the clustering pattern simply evolves as the galaxies follow the
continuity equation without losing their identity.  It can be shown
(Nusser \& Davis, 1994; Fry, 1996) that bias for such galaxies evolves
as
\begin{eqnarray}
b(z)=1+\frac{b_0-1}{D(z)},
\end{eqnarray}
where $D(z)$ is the linear growth rate for clustering, and $b_0$ is
the bias at the present epoch, $b_0=\sigma_{8,RG}/\sigma_{8,m}$, where
$\sigma_8$ is the rms fluctuation amplitude in a sphere of radius 8
h$^{-1}$Mpc.  We calculate $\sigma_{8,RG}$ assuming a correlation
radius $r_0\sim 10$ h$^{-1}$Mpc for radio galaxies (Peacock \&
Nicholson, 1991). We calculate $\sigma_{8,m}$ using a COBE-normalized
$\Lambda$CDM model with $\Omega_0=0.4$, $h_0=0.65$, $\Lambda=0.6$.
The resulting prediction for $w_{th}(\theta)$ is presented in Figure
\ref{fig:wgals}. The match with the data is extremely good, especially
when one considers that there are no free parameters in the model. This
gives us confidence on our understanding of the processes regulating
clustering in general and in particular the one of radio sources, at
least in the nearby universe.

\section{Conclusions} 
We have presented an analysis of the properties of optical
counterparts of radio sources at the 1 mJy level. The optical
identifications have been obtained by matching together objects from
the FIRST and APM surveys over the region of the sky
$148.13^\circ\simlt {\rm RA(2000)}\simlt 218.13^\circ$,
$-2.77^\circ\simlt {\rm Dec} \simlt 2.25^\circ$. We found 4075 radio
objects with fluxes greater than 1~mJy to be identified in the APM
catalogue for $b_J \simlt 22$ within a matching radius of 2
arcsecs (id-sample). This corresponds to 16.7 per cent of the original radio
sample.  A cut for $b_J$ = 21.5 has subsequently been applied to
ensure a uniform optical sample: 3176 identifications are found
brighter than this magnitude limit.

Such a wide area sample allows detailed statistical studies on the
photometry of radio sources. Furthermore, with the application of the
methods introduced in Maddox et al., (1990a), we managed to divide
radio sources with optical identifications and $b_J \le 20.5$ into
radio galaxies and stellar objects (mainly QSO with some contamination
from nearby stars).  We find the population of radio galaxies to be
almost exclusively dominated by early-type galaxies, with
radio-to-optical ratios in the blue band between $10^2$ and $10^4$ and
very red colours (up to $b_J-R\sim 4$). Starbursting objects 
make up only a negligible fraction of the sample. QSO tend to have
$r_B\simgt 10^4$ and preferentially lie in the region $0\simlt b_J
 -R \simlt 1$ of the colour-radio flux diagram at all radio
fluxes.

Since passive radio galaxies are known to be reliable standard
candles, measurements of the apparent magnitude in the red band allowed us 
to divide the id-sample into two low-redshift and high-redshift 
catalogues, the first one - roughly complete to $z\simeq 0.3$ - 
made by radio galaxies and the second one including high-z QSO.

The completeness and homogeneity and defined redshift cut of the galaxy sample
allowed a direct comparison between models for the angular correlation
function and the data. We find an extremely good match between observational 
results and models if we treat radio galaxies as highly biased tracers of the
underlying dark matter distribution and if we assume a bias evolution
with look-back time of the form introduced by Fry (1996). This is in
agreement with previous results obtained by Magliocchetti et
al. (1999) for the whole sample of FIRST radio sources.

Note that, as well as the results presented in this Paper, such a
large sample of radio sources with optical identifications forms an
excellent basis for spectroscopic follow-up observations, which could,
amongst others, fully determine the form of the redshift distribution
of these objects at the mJy level out $z\simlt 0.3$. We will tackle
this issue in a future Paper where we will present results from the
2dF survey.
 
\section*{ACKNOWLEDGMENTS}   
We would like to thank Mike Irwin for the help provided during the
analysis of the APM data and Gianfranco de Zotti and Annalisa Celotti for very 
useful discussions and comments on the manuscript.


\begin{thebibliography}{}    
\bibitem[\protect\citename{Baugh et al. }1999]{Ba}
Baugh C.M., Benson A.J., Cole S., Frenk C.S., Lacey C.G., 1999, MNRAS, 305, 21
\bibitem[\protect\citename{Becker et al. }1995]{Be}
Becker R.H., White R.L., Helfand D.J., 1995, {ApJ}, {450}, 559 
\bibitem[\protect\citename{Bock et al. }1999]{Bo}
Bock D. C-J., Large M.I., Sadler, E.M., 1999, AJ, 117, 1578
\bibitem[\protect\citename{Bunn and White} 1997]{Bunn} Bunn E.F.,
White M., 1997, ApJ, 480, 6
\bibitem[\protect\citename{Colless }1999]{Col}
Colless M., 1999, Phil Trans R Soc Lond. A, 357, 105
\bibitem[\protect\citename{Condon et al. }1998]{Co}
Condon J.J., Cotton W.D., Greisen E.W., Yin Q.F., Perley R.A., Taylor G.B., 
Broderick J.J., 1998, AJ, 115, 1693
\bibitem[\protect\citename{Cress et al. }1996]{Cr}
Cress C.M., Helfand D.J., Becker R.H., Gregg M.D., White R.L., 1996,
{ApJ}, {473}, 7  
  \bibitem[\protect\citename{Fry} 1996]{Fry} Fry J.N., 1996, ApJ, 461,
L65 
\bibitem[\protect\citename{Georgakakis et al. }1999]{Geo}
Georgakakis A., Mobasher B., Cram L., Hopkins A., Lidman C,
Rowan-Robinson M., 1999, MNRAS, 306, 708
\bibitem[\protect\citename{Gruppioni et al. }1998]{grup}
Gruppioni C., Mignoli M., Zamorani G., 1998, { MNRAS}, { 304}, 199
\bibitem[\protect\citename{Hamilton }1993]{Ha1}
Hamilton A.J.S., 1993, {ApJ}, {417}, 19   
\bibitem[\protect\citename{Helfand }1997]{He}
 Helfand D.J. at al., 1997, AAS, 190, 4304
\bibitem[\protect\citename{hine} 1979]{Hine}
Hine R.G., Longair M.S., 1979, MNRAS, 188, 111 
\bibitem[\protect\citename{Hopkins et al. }1998]{Hop}
Hopkins A., Mobasher B., Cram L., Rowan-Robinson M. 1998, 
 MNRAS, 296, 839  
   \bibitem[\protect\citename{Loan et al. }1997]{Lo}
Loan A.J., Wall J.V., Lahav O., 1997, {MNRAS}, {286}, 994
   \bibitem[\protect\citename{Loveday et al. }1995]{Love}
Loveday J., Maddox S.J., Efstathiou G., Peterson B.A., 1995, ApJ, 442,457
   \bibitem[\protect\citename{Lu et al. }1996]{Lu}
Lu N.Y., Lyle Hoffman G.L., Salpeter E.E., Houck J.R., 1996, ApJS, 103, 331
\bibitem[\protect\citename{Maddox et al. }1990c]{Madc}
Maddox S.J., Efstathiou G., Sutherland W.J., Loveday J., 1990c, {
  MNRAS}, {242}, 43p        
\bibitem[\protect\citename{Maddox et al. }1990a]{Mada}
Maddox S.J., Efstathiou G., Sutherland W.J., Loveday J., 1990a, {
  MNRAS}, {243}, 692        
\bibitem[\protect\citename{Maddox et al. }1990b]{Madb}
Maddox S.J., Efstathiou G., Sutherland W.J., 1990b, {
  MNRAS}, {246}, 433    
\bibitem[\protect\citename{Maddox et al. }1996]{Mad}
Maddox S.J., Efstathiou G., Sutherland W.J., 1996, {
  MNRAS}, {283}, 1227    
\bibitem[\protect\citename{Magliocchetti et al. }1998]{Mag}
Magliocchetti M., Maddox S.J., Lahav O., Wall J.V., 1998, {MNRAS}, {
  300}, 257
\bibitem[\protect\citename{Magliocchetti et al. }1999]{Maga}
Magliocchetti M., Maddox S.J., Lahav O., Wall J.V., 1999, {MNRAS},
309, 943
\bibitem[\protect\citename{Magliocchetti et al. }2000a]{Manu1}
Magliocchetti M., Bagla J., Maddox  S.J., Lahav O., 2000a, MNRAS, 
314, 546
\bibitem[\protect\citename{Magliocchetti et al. }2000]{Manu}
Magliocchetti M., Maddox  S.J., Wall  J.V., Benn C.R., Cotter G., 2000,
MNRAS, 318, 1047  [MA2000]
\bibitem[\protect\citename{Magliocchetti et al. }2001]{Manu2}
 Magliocchetti M., Celotti A., Danese L., 2001, MNRAS, submitted
\bibitem[\protect\citename{Martini et al. }2001]{Martini}
Martini P., Weinberg D.H., 2001, ApJ, 547, 12
\bibitem[\protect\citename{Masci et al. }2001]{Masci}
Masci F.J., Condon J.J., Barlow T.A., Lonsdale C.J., Xu C., Shupe D.L., 
Pevunova O., Cutri R., 2001, PASP, 113, 10
\bibitem[\protect\citename{Matarrese et al.} 1997]{Mat} Matarrese S.,
Coles P., Lucchin F., Moscardini L., 1997, MNRAS, 286, 115  
\bibitem[\protect\citename{McLure et al.} 1999]{Mc}
McLure R.J., Kukula M.J., Dunlop J.S., Baum S.A., O'Dea C.P., Hughes D.H., 
1999, MNRAS, 308, 377
    \bibitem[\protect\citename{Mo and White} 1996]{Mo} Mo H., White
S.D.M., 1996, MNRAS, 282, 347     
\bibitem[\protect\citename{Moscardini et al.} 1998]{Mos}
Moscardini L., Coles P., Lucchin F., Matarrese S., 1998, MNRAS,
299, 95     
\bibitem[\protect\citename{Nusser and Davis} 1994]{Nu} Nusser A.,
Davis M., 1994, ApJ, 421, L1    
\bibitem[\protect\citename{Pea }1991]{Pea2}
Peacock J.A., Nicholson D., 1991, {MNRAS}, {253}, 307
\bibitem[\protect\citename{Peacock2} 1996]{peacock2}
Peacock J.A., Dodds S.J., 1996, MNRAS, 267, 1020   
\bibitem[\protect\citename{Peebles }1980]{Pe}
Peebles P.J.E., 1980, {The Large-Scale Structure of the Universe},
Princeton University Press   
\bibitem[\protect\citename{Rixon et al. }1997]{Rix}
Rixon G.T., Wall J.V., Benn C.R., 1991, {MNRAS}, {251}, 243 
\bibitem[\protect\citename{Sadler et al. }1999]{Sad}
Sadler E.M., McIntyre V.J., Jackson C.A., Cannon R.D., 1999, PASA, 16, 247   
\bibitem[\protect\citename{Urry }1995]{urry}
Urry C.M., Padovani P., 1995, { PASP}, { 107}, 803   
\bibitem[\protect\citename{White et al. } 1997]{Whi}
White R.L., Becker R.H., Helfand D.J., Gregg M.D., 1997, {ApJ},
{475}, 479                                                       
\end{thebibliography}
\end{document}